\documentclass{article}
\input{tcilatex}

\begin{document}

\section{On the completeness of quantum mechanics$^{\dagger }$}

\begin{center}
\textbf{M.Kupczynski\medskip }

\ Department \ of Mathematics and Statistics , Ottawa University\medskip
\end{center}

\begin{quotation}
\textbf{Abstract.} Quantum cryptography, quantum computer project,
space-time quantization program and recent computer experiments reported by
Accardi and his collaborators show the importance and actuality of the
discussion of the completeness of quantum mechanics (QM) started by Einstein
more than 70 years ago. Many years ago we pointed out that the violation of
Bell's inequalities is neither a proof of completeness of QM nor an
indication of the violation of Einsteinian causality. We also indicated how
and in what sense a completeness of QM might be tested with the help of
statistical nonparametric purity tests. In this paper we review and refine
our arguments. We also point out that the statistical predictions of QM for
two-particle correlation experiments do not give any deterministic
prediction for a single pair. After beam is separated we obtain two beams
moving in opposite directions. If the coincidence is reported it is only
after the beams had interacted with corresponding measuring devices and two
particles had been detected. This fact has implications for quantum
cryptography. Namely a series of the measurements performed on the beam by
Bob and converted into a string of bits (secret key) will in general differ,
due to lack of strict anti-correlations , from a secret key found by Alice
using the same procedure.
\end{quotation}

\bigskip

\textbf{Keywords:} Bell's inequalities, completeness of quantum mechanics, \
purity and ergodicity tests, \ EPR correlations, quantum cryptography

\textbf{PACS Numbers}: 03.65. Bz, 03.67. -a, 03.67. Dd, 03.67.Hk

$^{\dagger }$ Extended version of a paper presented at the International
Conference on Theoretical Physics,Paris, UNESCO, July 22nd-27th,2002

\subsection{1.Introduction}

\noindent Let us imagine that we are sitting on a shore of an island on a
lake watching a sunset. We see the birds flying , the leaves and branches
are moving with a wind, a passage of a boat produces all interesting
patterns on the surface of the water and finally we hear regular waves
hitting the shore. Finally a big round circle of the sun is hiding under the
horizon leaving a place for beautifully illuminated clouds and later for the
planets and stars. All these physical phenomena are perceived by \ us in
three dimensions and they are changing in time usually in the irreversible
way.

To do the physics we have to construct mathematical models leading to the
predictions concerning our observations and measurements this is why we
created concepts of material points, waves and fields. For Newton light was
a stream of small particles for Maxwell light was an electromagnetic wave
moving in a continuous invisible medium called ether , similarly to waves on
a water. With abandon of the concept of ether in special theory of
relativity the image of the propagation of light became less intuitive. A
discovery of \ the fact that the exchanges of the energy and of the linear
momentum between light and matter are quantized gave a temptation to
represent again the light as a stream of indivisible photons moving
rectilinearly and being deflected only on the material obstacles or absorbed
and emitted by the atoms. This picture together with an assumption that each
indivisible photon may pass only by one slit or another and that the
interaction with a slit through which it is passing does not depend on a
fact that the other slit is open or closed is clearly inconsistent with the\
observed interference pattern. Anyway photons are not localizable objects
but the same argument could be repeated for a double slit electron
experiment. Therefore we discover that the light and the matter may present
wave and corpuscular behavior in the mutually exclusive ( complementary )
experimental arrangements.

Moreover there is the wholeness in the experiment : a source is prepared and
calibrated, it interacts with the experimental arrangement and the modified
source and/or the final numerical results the measurements are found. The
only picture given is a black box picture. As an input we have an initial
''beam'' entering a box as an output we have a modified ''beam'' (
''beams'') or a set of counts of various detectors. Quantum mechanics (QM)
does not give any intuitive spatio-temporal picture of what is physically
happening in the box. The QM gives only the predictions about the final
''beams'' and about statistical distribution of the counts of the detectors.
Let cite Bohr$\left[ 1\right] $ :..\ ''Strictly speaking , the mathematical
formalism of quantum mechanics and electrodynamics merely offers rules of
calculation for the deduction of expectations \ pertaining to observations
obtained under well-defined experimental conditions specified by classical
physical concepts''. This statement is valid not only for the description of
standard atomic phenomena but also for the S- matrix description of all
scattering processes of elementary particles and for the stochastic models
describing the time evolution of trapped molecules, atoms or ions. The
quantum mechanics and new stochastic approaches have no deterministic
prediction for a single measurement or for a single time -series of events
observed for a trapped ultracold atom. The predictions being of statistical
or of stochastic character apply to the statistical distribution of the
results obtained in long runs or in several repetitions of the experiment.
We will give a careful epistemological discussion of the experiments with
trapped atoms, quantum dots and qubits in the subsequent paper. In this
paper we will limit our discussion to standard experiments and to standard
QM.

For example in a two slit interference experiment with low intensity source
of monochromatic light we can ''measure'' its intensity by the counts
registered by a photon detector , we can control the intensity of the source
by opening and closing regularly a collimator in order to send regular
pulses of light. In this case we estimate an average intensity of the
''beam'' \ \ \ \ \ \ (number of clicks of the detector interpreted as a
number of photons absorbed) . If our screen behind the slits is in the form
of a panel of photon detectors, after waiting long enough, we find spatial
statistical distribution of the counts registered by detectors consistent
with the classical interference pattern. However we are not allowed to
imagine a light as a beam of somewhat localized and separated photons moving
rectilinearly hitting one after another a double slit screen, passing by
only one of the slits and after continuing their way to the detectors.

Similarly if we return to a passage of the boat on a lake we can detect and
even measure the energy and the momentum transferred by the regular waves
hitting a buoy close to the shore. We could even tell that we observe a beam
of ''wavelons'' hitting the shore. There would be no comparable transfer of
the energy and momentum on any buoy on a deep water away from a shore.
Therefore we could not make an image of the boat producing a beam of
wavelons which after rectilinear propagation hit the shore. Of course we can
see changes on the surface of \ the considerable portion of a lake but in
quantum physics we do not see the'' lake''.

This example shows a danger of image making. Wrong images lead to
contradictions and to wrong deductions. A classical mechanics also
concentrates on the description of the observations without creating too
many images. The Sun and the Earth are represented mathematically by
material points characterized at each moment of time by their masses,
positions and velocities. If in some inertial frame the initial positions
and velocities are known Newton's equations allow us to determine a
subsequent motion of these points which agrees \ remarkably well with a real
motion of Earth around the Sun. There is no speculation by what mechanism a
change in the position of one body causes an instantaneous change in the
acceleration of its far away partner but it does not harm a success of the
model. Of course a quest for more detailed understanding of the mutual
interactions between far apart bodies led to the progress in physics namely
to the development of classical electrodynamics and to the creation of
general theory of relativity.

In spite of the fact that the QM gives only statistical predictions on the
outcomes of the various experiments a claim is made that QM gives a complete
description of the physical phenomena and even the most complete description
of the individual system. Einstein has never accepted this claim and in his
famous paper written with Rosen and Podolsky$\left[ 2\right] $ about EPR\
paradox he started a fruitful discussion on the completeness of \ QM and on
general epistemological foundations of physics. This discussion continues
till now.

Many physicists adopt the statistical interpretation of QM$\left[ 3\right] $
in which a wave function describes only an ensemble of identically prepared
physical systems and the wave function reduction is a passage from the
description of the whole ensemble of these systems to the description of a
sub-ensemble satisfying some additional conditions. The statistical
interpretation is free of paradoxes because a single measurement does not
produce the instantaneous reduction of the wave function. The statistical
interpretation leaves a place for the introduction of the supplementary
parameters (called often hidden variables) which would determine the
behavior of each particular physical system during the experiment. Several
theories with supplementary parameters (TSP) have been discussed $\left[ 4%
\right] $ .

The QM gives predictions for spin polarization correlation experiments(SPCE)
dealing with pairs of electrons or photons produced in a singlet state. In
order to explain these long range correlations Bell$\left[ 5\right] $
analyzed a large family of TSP so called local or realistic hidden variable
theories (LRHV) and showed that their predictions must violate,\ for some
configurations of the experimental set-up, the quantum mechanical
predictions. Bell's argument was put into experimentally verifiable form by
Clauser, Horne,Shimony and Holt$\left[ 6\right] $. Several experiments in
particular those by Aspect et al. $\left[ 7\right] $ confirmed the
predictions of quantum mechanics. Many physicists concluded that if a TSP
wants to explain the experimental data it must allow for the faster than
light communication between particles and violate Einstein's separability.
Even without deep reasoning one can see that this conclusion must be flawed.
Let us imagine a \ huge volcanic eruption taking place somewhere in the
middle of the Pacific Ocean, the sunami waves hitting the shores of Japan
and America will be correlated in a natural way. Long range correlations
come from the memory of the past events and time evolution and they have
nothing to with extra luminal communications between far away objects.

It was shown by many authors that the assumptions made in LRHV are more
restrictive that they seem to be and the Bell's inequalities may be violated
not only by quantum experiments but also by macroscopic ones. Let us mention
few of them. Accardi gave an extensive discussion of non-Kolmogorovian
character of the quantum probabilities$\left[ 8\right] $ and noticed that
the most important assumption needed to prove the Bell inequalities is not a
locality assumption but the use of the same probability space. Pitovsky
constructed local hidden variable model $\left[ 9\right] $ which could
reproduce the quantum mechanical polarization predictions. Aerts$\left[ 10%
\right] $ showed that non-Kolmogorovian character of the quantum
probabilities is due to the indeterminacy on the measurements in contrast to
the indeterrminacy on the classical states De Baere$\left[ 11\right] $
strongly claimed that the violation of Bell inequalities is due to
non-reproducibility of a set of hidden variables in the subsequent
experiments.

In 1976 we noticed that if one associates to each EPR pair a couple of
bi-valued spin functions S$_{1}$(a) and S$_{2}$(b) where a and b are the
unit direction vectors of polarizers it is not clear how we can use the
integration over the finite dimensional space of hidden variables to
describe all these random experiments. Besides we noticed that we can not
prove rigorously the Bell inequalities for the empirical spin expectation
functions because in the runs of the different experiments the sets of
couples of spin functions may be different. Nevertheless it seemed plausible
to us that after averaging the approximate Bell's inequalities would still
be valid. We communicated our comments to Bell during our short stay in
Geneva in September 1976 but we did not publish them.

In 1982 Bell brought to our attention the Pitovsky's paper. The model was
using axiom of choice and was quite difficult to understand but it was able
to reproduce QM predictions for the SPCE. We noticed that the model can be
simplified and that by using the particle beams described by Pitovsky's
particular spin functions one could reproduce quantum mechanical predictions
and avoid Bell's inequalities $\left[ 12,13\right] $. We noticed also$\left[
14\right] $ that in all proofs of Bell's theorem $\left[ 15\right] $ one is
using (directly or indirectly) the assumption that in the moment of
production both members of each pair of quanta have unknown but well defined
and strictly correlated spin projection values in all directions,
distributed according to some joint and unknown probability distribution,
and if we try to measure a spin projection in a particular direction a
measuring device can register a correct value or fail to register it with a
small probability. Only in this case one can obtain predictions for all
different random experiments (A,B) where A denotes polarizer used for a
particle 1 and B a polarizer used for a particle 2 by conditionalization
from a single sample (probability) space (at the time being we did not know
the Accardi paper$\left[ 8\right] $).

However the photons and the electrons are not small spinning balls and it is
well known that the QM is a contextual theory. Namely a value of a physical
observable, here a spin projection, associated with a pure quantum ensemble
and in this way with an individual physical system , is not an attribute of
the system revealed by a measuring apparatus; it turns out to be a
characteristic of this ensemble created by its interaction with the
measuring device . It is therefore meaningless to consider joint spin
projection distributions in all directions and the quantum mechanical
predictions can not be hoped to be reproduced from the TSP models of this
type. In the modified Pitovsky model$\left[ 13\right] $ a quantum has a spin
up in \textbf{A} direction if it is a member of an ensemble of particles
transmitted by a polarizer A. The spin functions , in the model, describing
interactions of the quanta with the polarizers, have well defined values on
all unit vectors on a sphere but a passage through a given polarizer depends
on the probability distribution of the unit vectors representing this
polarizer , distributed statistically around a macroscopic orientation
vector \textbf{A}. For this reason there is no deterministic prediction on
the behavior of two members of each EPR. pair and strict anti-correlations
may not be anticipated before being observed. Similar conclusions were
formulated by Schroeck$\left[ 16\right] $ who analyzed the EPR\ experiment
using the measurement scheme of stochastic quantum mechanics$\left[ 17\right]
$.

In$\left[ 14\right] $ we\ recalled the paradox of Bertrand $\left[ 18\right] 
$who clearly demonstrated the importance of a direct link of the
probabilistic model with a random experiment which it wants to describe. We
underlined that the different experiments are described by the probability
density distributions defined on their own probability spaces and they can
be described by conditionalization from a single probability space only if
all of then can be performed simultaneously on each member of a statistical
population.

In view of these observations we had no doubt that the violation of Bell
inequalities did not mean that if a TSP wanted to explain the EPR or other
quantum mechanical experiments it had to violate Einstein's causality.
Because \ we were quite satisfied with the statistical description of \ the
phenomena and we were not interested in inventing a new TSP\ we stopped
working on the subject.

With the advent of quantum cryptography introduced by Bennet and Brassard $%
\left[ 19,20\right] $ streams of photons were proposed to be used to
transmit a secret key and the fact that any measurement affects the quantum
state could be used to detect eavesdropping. We find questionable the use of
single photons but certainly a scheme is realizable by using the pulses of
the light polarized in a particular fashion. There is a bigger problem with
a model by Ekert $\left[ 21\right] $ in which strictly correlated EPR pairs
are used in order to transmit the same secret key to Alice and Bob and the
Bell's theorem is used as a test for eavesdropping. According to us there is
no strict anti-correlation on the individual level so the argument does not
hold.

In 2001 we received a preprint of Accardi and Regoli$\left[ 22\right] $ in
which they presented the results of the computer experiment violating Bell's
inequalities and gave many other arguments and references showing that there
was no contradiction between quantum theory and locality. According to
Accardi a violation of the Bell's theorem is due to the concept of chameleon
effect $\left[ 23\right] $( the dynamical evolution of the system depends on
the observable \ one is going to measure) which is closely related to the
fact that the QM is a contextual theory. Accardi's biological comparison is
nice. In fact a quantum particle has no attributes by itself. A quantum
particle shows different behavior in the interactions with different
experimental arrangements similarly to a chameleon whose color depends
whether he is sitting on a leaf or on a trunk of a tree.

The tests of Bell's theorem led to many beautiful experiments$\left[ 24%
\right] $\ . However it seems that the epistemological implications of the
demonstrated violation of Bell's inequalities are still not fully understood
by the majority of the physicists $\left[ 25\right] $.

Moreover we realized that our contributions, still valid, to the subject
seem to be unknown , forgotten or not understood. In particular we
introduced direct tests of the completeness of QM$\left[ 12,26)\right] $ by
means of the purity tests which have been completely neglected.

This is why we decided in this paper to refine our old reasoning $\left[
12,13,14,26\right] $ and to add some new much simpler arguments.

\bigskip

\subsection{2.Completeness of a statistical theory.}

A statistical description is not a description of the objects but it is a
description of the regularities observed in large populations or in a series
of repeated random experiments performed with the objects. Let see it on
examples.

We consider a series of coin flipping experiments. Instead of coins having
head and tails we have coins with one side ''blue''(B) and one side
''red''(R). If we want to provide a complete description of a coin using the
concepts of classical physics and mechanics we may say that a coin is a
round disk of a given diameter etc.. We can find also its mass, volume,
moment of inertia etc. All these attributes (values of classical
observables) describe ''completely'' a coin from a classical point of view.
We have also at our disposal various coin- flipping devices:D1, D2,.. All of
them look from outside in the same way: you have a place to put a coin, \
one of the faces up, and a button to push on. A coin is projected and you
see it flying , rotating and finally it falls on the observation plate.

EXPERIMENT 1 (E1). We start with a device D1 and we use only one coin. At
first we do not pay attention what is a color of a face of the coin which we
put up. For example we record a series of outcomes:BBRBRRRB... At the first
sight it is a time series of events without any regularity. We decide now to
be more systematic and to put always a face B up into the device. To our big
satisfaction we obtain a simple time series: :RRRRRR... If instead we put a
face R up into device we obtain: BBBBB.... From an empirical \ point of view
our description of the phenomenon is complete. A device D1 is a classical
deterministic device such that if we insert into it a particular coin it
changes a face B up of the coin into a face R up of the coin and vice versa.

However we do not see only the final result we see also the coin
flying,revolving and landing. If you are a physicist you would like to
understand why so complicated phenomenon gives a simple deterministic
result. Let us imagine that we are allowed to see the interior of the
device. If we see that D1 gives always to the coin the same initial linear
velocity and the same initial angular velocity then knowing the laws of
classical mechanics and taking into consideration air resistance but
neglecting the influence of the air turbulences, caused by the revolving
coin, \ we can,with a help of a computer , reproduce a flight of the coin
and deduce that the coin placed with one face up will land always on the
observation plate with another face up. It would provide a complete
description of the phenomenon. Even if we were unable to make calculations
we could anticipate a result and we could say that we understood
''completely'' a phenomenon. Of course we took the Newton's equations for
granted but in some point looking for the explanation we have to stop asking
a question: ''Why?''.

EXPERIMENT 2 (E2). We take the same coin and a device D2. On a basis of the
previous experiment we start by placing the coin always with face B up and \
we preform several series of trials. To our surprise we get a time series of
results BRBRBRB...or \ RBRBR...We obtain similar results if we place the
face R up. A complete empirical explanation of the phenomenon is that D2
produces completely deterministic alternating series of outcomes. The only
uncertainty is a first result. It shows that a device has some memory. For
example a flipping mechanism of the D2 can be identical to the flipping
mechanism of the D1 with one difference that the inserted coins are rotated
around a horizontal axis before being flipped\ with a rotating mechanism
keeping a memory of events : each 180$%
%TCIMACRO{\UNICODE[m]{0xb0}}%
%BeginExpansion
{{}^\circ}%
%EndExpansion
$ rotation is followed by 360$%
%TCIMACRO{\UNICODE[m]{0xb0}}%
%BeginExpansion
{{}^\circ}%
%EndExpansion
$ and vice versa. To understand ''completely'' the phenomenon we look in the
interior of the device and we repeat the analysis we did for D1.

EXPERIMENT 3 (E3) We replace the device D2 by a device D3. We repeat several
times the experiment with the face B up and after with the face R up. We
obtain various time series which seem to be completely random. We call a
colleague statistician for help. He checks that the observed time series is
random. He observes that relative empirical frequencies of observing the
face B in long runs are close to 0.5. It concludes that each experiment is a
Bernoulli trial with a probability p=0.5. Using this assumption he can make
predictions concerning the number of faces B observed in N-repetitions of
the experiment and compare them with the data. A statistical description of
the observed time-series of results is complete and it may be resumed in the
following rigorous way: Anytime we place the coin into the device D3 there
are two outcomes possible each obtained with \ a probability 0.5. A
probability 0.5 it is not the information about the coin. It is not the
information about the device D3. It is only the information about the
statistical distribution of outcomes of random experiments : inserting the
coin into the device, pushing the button on and observing the result. This
is why a statement: the coin, if flipped, has a probability 0.5 to land with
the face B up is incorrect. All devices considered above are flipping
devices but the statistical distribution of the results they produce are
different. We could correct this statement by adding: if flipped with the
device D3 , but one has to remember what does it mean. Once again to
understand completely the phenomenon we could look in the interior of the
device D3 .For example we might find that D3 is identical to the device D1
but before flipping there was some mechanism rotating the coin in a random
(pseudo random) way. It would allow us to understand more ''completely'' the
phenomenon but it would not give us any additional information about the
statistical distribution of results.

There could be however an advantage of this ''complete'' description of the
phenomena. Let us imagine that to each device considered above we add a
ventilator blowing on the coin when it is flying. It would certainly modify
statistical distribution of results in the experiments E1 and E2. From the
empirical point of view the device D1 with a ventilator it is a new device
D'1 so we have a new random experiment and a new statistical distribution to
be found. However on this level we are unable to predict how this new
description originates from a previous one. On the contrary a knowledge of
the ''complete '' description of the phenomenon describing a flight of the
coin produced by D1 could be used to predict the modifications induced by
the wind produced by the ventilator. If we had a classical theory describing
time evolution of the air turbulences and its interactions with the coin (
which we don't have) we could in principle determine the possible
trajectories of the coin and deduce the changes in the statistical
distribution of experimental results.

In all these experiments we saw the coin flying and we could look inside the
experimental devices. If we did not have this knowledge but only the
knowledge of final results the only unambiguous description would be a
statistical one. Probably we could invent infinite number of ''
microscopic'' \ hidden variable models agreeing with observation but we
would not gain any better understanding of the results. This resembles to
the situation in quantum mechanics. We have a stable source producing some
beam. We place some detector in front of the beam which clicks regularly
what makes us believe that we have a beam of \ some invisible ''particles''
\ having some constant intensity. We take the detector out and we pass our
beam by the experimental arrangement ( a device) and we observe a time--
series of the possible final outcomes. QM gives us the algorithms to
calculate the time independent probability distributions of the outcomes
giving no information how the time -series is \ building up. Einstein
understood very well the statistical description of the experiments given by
QM but he believed that this statistical description should be ''completed
'' by some''microscopic'' description explaining how the observed time-
series of \ the results is building up. It seems to us that if such
description existed, it would be extremely complicated and not unique so
perhaps not very useful.

Even if one does not think that such ''microscopic '' description is needed
a hypothesis, that such description is possible, suggests that there is some
information in the time- series of the results not accounted for by the
statistical description given by QM. If it was true a careful analysis of
time- series of results could reveal some structure not explained by QM what
would imply that statistical description provides the incomplete statistical
description of the data.

Therefore a question whether a particular statistical description of the
phenomenon is complete or not, it is an experimental question which can be
asked and answered independently of the existence of a ''microscopic''
description of the phenomenon. The answer can be obtained with the help of
the purity tests which were proposed many years ago$\left[ 12,26\right] $%
\textbf{\ }and which we will discuss in some detail in the moment. Let us
continue a discussion of simple macroscopic experiments which hopefully will
help to understand our point of view.

As we saw in the experiment E1 any time- series obtained could be
interpreted as a series of the results of the consecutive repetitions of the
identical Bernoulli trials each characterized completely by a probability p
= 0.5. Let us consider now another random experiment.

EXPERIMENT 4 (E4) .There is a box containing the coins but we do not see
what is in the box. There are 51 blue coins and 51 red coins in the box
.With closed eyes we mix well the coins , we draw one coin from the box, we
place it on a table and finally we open the eyes and we record the color of
the coin. We decide to repeat this experiment n-times up to n=100 and after
100 repetitions we return all coins drawn into the box without looking at
them.. If we use the various runs(samples) with n bigger for example than 95
to estimate the frequency of observing a blue face we will find that it is
close to 0.5 so we are ready to conclude that a probability of drawing a
blue face in each draw is equal to 0.5 and mathematically speaking the
experiment E4 is the same as the experiment E3. Our friend statistician
tells us not to jump into conclusion too fast because if initially in the
box we have 2N coins ( N red and N blue) on the average we find 50\% of blue
coins in a sample but a time- series is different and in principle we can
discover it by more detailed statistical analysis of this series. In the
case of the Bernoulli trials at each repetition the probability of obtaining
B is the same. On the contrary the probability of drawing a blue coin in the
k-th draw depends on a number of blue coins drawn already. Namely if among
the k draws there were m blue coins in this case the probability of
obtaining a blue coin in the next draw is \ \ \ \ \ \ \ \ \ \ \ \ \ \ \ \ \
\ \ \ \ \ \ \ \ \ \ \ \ \ \ \ \ \ p(k+1)=p(k+1,m, N)= (N-m)/(2N-k).
Consequently we will have a hypergeometric distribution instead of binomial
etc.

Thus in the E4 we have a succession of the different dependent random
experiments when in the E3 we have a succession of the identical independent
random experiments. The averages of two time series are consistent but they
have completely different structure. In this case a statement that a
probability of drawing a blue face is in each draw equal to 0.5 is not only
incomplete but it is also incorrect. If we modify the experiment E4 namely
by returning the coin to the box after each draw our new experiment is, for
samples of a size smaller than 102, completely equivalent to the experiment
E3. On the ''microscopic '' level there is however one fundamental
difference: in the E4 the coins in the box are always either blue or red
when the coin in the experiment E3 is neither blue nor red but unfortunately
this difference is not seen from the existing data. To see more easily how
such ''microscopic'' differences could be detected by performing additional
experiments we will discuss another macroscopic experiments with the coins.

EXPERIMENT\ 5 (E5) There is a box ,which contains now\ 50 blue coins and 50
red coins having all other physical properties identical. A button is pushed
and a mechanical arm picks at random one of the coins in a box and inserts
it into the device D3. The result B or R is recorded and handed to the
experimenter and a coin is returned to the box.

EXPERIMENT 6 (E6) The only difference with E5 is that instead of containing
50 red and 50 blue coins a box contains 100 two-sided coins identical to the
coin \ used in the experiments E1-E3.\ All other physical properties of
two-sided coins are the same as the physical properties of the coins in the
E5.

These two experiments produce random time-series of results which do not
allow to find any significant difference between them. Two physicists agree
with this statement but they cannot agree how to interpret the results. One
of them ,a pupil of Einstein, says : we have a statistical mixture(mixed
statistical ensemble) in the box of the same number of blue and red coins
and because we draw the coins from the box with replacement thus on average
we observe 50\% of blue coins in each run of the experiment. A second ,
pupil of Bohr , says: it is a nonsense we have simply a pure statistical
ensembles of quantum coins each in the same pure quantum state, such that
each of them has simply a probability 0.5 to become blue or red after
interacting with the measuring device.

They meet a statistician who confirms that the experiments give the
indistinguishable results and tells them that without performing
supplementary experiments one cannot decide whose model is a correct one. He
tells them that in a mixed statistical ensemble every of it sub-ensembles
can in principle have different observable statistical properties. On the
contrary if one has a pure statistical ensemble all of its sub-ensembles
have the same properties as the initial ensemble.

Our physicists agree with the statistician and they make a hole in the boxes
containing coins and they decide to remove the same number of coins in the
E5 and in the E6 before proceeding with several repetitions of their
experiment.

If they removed by chance the equal number of blue and red coins in the E5
no difference could be noticed but if they by chance changed a proportion of
blue coins in their box they could see the difference in long runs of the
experiment. If less coins were left in the box the differences could be
bigger. For example with 4 blue and 6 red coins in the box the probability
of B became 0.4 instead of 0.5. With one blue coin in a box they would get
p=1.

Following the same protocol for the experiment E6 they did not register any
significant difference in the results. They concluded that there is a
''microscopic'' difference between the E5 and the E6 and that Einstein's
model apply to the E5 and Bohr's model apply to E6.

Therefore a claim that the QM\ gives a complete description of the
individual quantum system may not be disproved by any philosophical argument
nor by a mathematical theorem it may only be disproved by the experimental
data. The only situation when the statistical predictions on the results of
the experiments performed on the ensemble of identically prepared individual
systems can be said to describe completely the interaction of the individual
system with the experimental device occurs when the observed time-series is
random and the statistical ensemble is pure.

The assumption of the completeness of the statistical description provided
by QM is not only unnecessary but it is counter- productive. The
experimentalists are interested only in testing the statistical
distributions of the experimental results in long runs without even trying
to analyze the observed time-series. They eliminate the ''bad'' experimental
runs, sometimes without finding any logical reason for doing it, simply
because in the theory there is no place for them. We have enormous amount of
data accumulated. If we performed the tests of \ the randomness and the
purity tests$\left[ 26\right] $ on these data perhaps we would discover new
statistical regularities in the time- series we had never thought of. Let us
talk now about the origin of the purity tests.

\subsection{3.Purity tests}

The QM did recognize that the purity of the quantum ensemble is important if
a claim was made that QM provides the complete description of the individual
system. However to define the purity the QM concentrated on the preparation
stage of the experiment. Namely a system was said to be in a pure quantum
state if it passed by a maximal filter or if a complete set of commuting
observables was measured on the system.

A purity of the statistical ensemble describing the interactions of such
prepared state with some other experimental device was not considered.
Moreover it was not clear how \ we could know that a filter is maximal and
how do we construct it. Besides in the axiomatic quantum mechanics,
initiated by a paper by Birkhoff and von Neumann$\left[ 27\right] $, \ it
was claimed that to each vector in the Hilbert space corresponds a
realizable physical state of a physical system and that the Hilbert space
description is general enough to describe all imaginable future phenomena.
The last claim was nicely refuted by Mielnik$\left[ 21\right] $ who showed
that one can imagine infinitely many non-Hilbertian ''quantum'' worlds.
Inspired by first two Mielnik's papers we decided to analyze various general
experimental set-ups which could be used to investigate the phenomena
characterizing the ensembles of particle-beams. We considered \ the sources
of some hypothetical particle beams, detectors( counters),
filters,transmitters and instruments This analysis$\left[ 29\right] $ led us
to the various conclusions. Let us list those which are pertinent to the
topic of this paper:

1) Properties of the beams depend on the properties of the devices and
vice-versa and are defined only in terms of the observed interactions
between them. For example a beam \ b is characterized by the statistical
distribution of outcomes obtained by passing by all the devices d$_{i}$. A
device d is defined by the statistical distributions of the results it
produces for all available beams b$_{i}$. All observables are contextual and
physical phenomena observed depend on the richness of the beams and of the
devices.

2) In different runs of the experiments we observe the beams b$_{k}$ each
characterized by its empirical probability distribution. Only if an ensemble
\ss\ of all these beams is a pure ensemble of pure beams we can associate
the estimated probability distributions of the results with the beams b$\in $%
\ss\ and with the individual particles members of these beams.

3) A general operational definition of a pure ensemble \ss\ :

A pure ensemble \ss\ of pure beams b is characterized by such probability
distribution s(r) which remains approximately unchanged:

(i) for the new ensembles \ss $_{i}$ obtained from the ensemble \ss\ by the
application of the i-th intensity reduction procedure on each beam b$\in $\ss

(ii) for all rich sub-ensembles of \ss\ chosen in a random way

In 1974 we noticed$\left[ 30,33,36\right] $ that if the initial two-particle
states in strong- interaction physics were mixed with respect to some
additional parameter,for example the impact parameter, and if we wrote the
unitary S-matrix as S=I$\oplus $S$_{1}$instead of $\ $S=iT+I then the
probability would be conserved but the optical theorem could not be proven.
Because all extrapolations to the forward direction were unreliable$\left[
35,36\right] $ therefore the only way to check our hypothesis was to find
this particular impurity of initial states in high energy scattering. . For
this purpose we proposed various purity tests$\left[ 31,32,34\right] $. The
experiments to test our hypothesis were never done. In 1984 we noticed $%
\left[ 26\right] $ that the similar purity tests could be used to test the
completeness of the quantum mechanics.

The idea is extremely simple and it was explained in the previous section:
one has to study in detail \ time-series of the experimental results and
look for some fine structure.

Besides one has to compare different runs of the same experiment looking for
statistically significant discrepancies. Namely if b$_{i}$ is a beam of m$%
_{i}$ particles produced by a stable source\ O in the time interval $\left[
t_{i},t_{i}+\Delta t\right] $ \ we obtain a sample S$_{i}$ by measuring an
observable $\gamma X$ on the beam b$_{i}$. We may also obtain the families
of the beams b$_{i}$(j), where j denotes j-th beam intensity reduction
procedure applied to the beam b$_{i}.$ Measuring $\gamma X$ on the beams b$%
_{i}$(j) we obtain new samples S$_{i}$(j).

To test the purity of the beams produced by O we test a hypothesis

H$_{0}$: All the samples S$_{i}$and S$_{i}$(j) \ are drawn from the same
unknown statistical population of the random variable X associated to the
observable $\gamma X.$

An extensive use of the non-parametric statistical tests is needed$\left[ 34%
\right] $. We are in 2002 nobody did the purity tests. Unfortunately all the
experiments confirming the validity of the quantum mechanical statistical
predictions and the violation of Bell's theorem eliminates LHRV\ but are
unable to prove the completeness of QM. Let us see this in more detail

.

\subsection{4.Bell's Theorem}

To each random experiment we associate a random variable X, a probability
space S and a probability density function f$_{X}$(x) for all x$\in $ S.

If\ X is a discrete random variable $\stackunder{x}{\sum }$ f$_{X}$(x)=1 and
P(X=x)=f$_{X}$(x) If X is a continuous random variable$\stackunder{S}{\int }$%
\ f$_{X}$(x)dx=1 and

P(a$\leq $X$\leq $b)\ =$\stackunder{a}{\int }$\ $^{b}$ f$_{X}$(x)dx \ \ \ \
\ \ \ \ \ \ \ \ \ \ \ \ \ \ \ \ \ \ \ \ \ \ \ \ \ \ \ \ \ \ \ \ \ \ \ \ \ \
\ \ \ \ \ \ \ \ \ \ \ \ \ \ \ \ \ \ \ \ \ \ \ \ \ \ \ \ \ \ \ \ \ \ (1)

where P(a$\leq $X$\leq $b) is a probability of finding a value of X included
between a and b. Note that P(X=x) = 0 for all x$\in $S\textbf{.}

If in a random experiment we can measure simultaneously values of k- random
variables\ X$_{1}$,...X$_{k}$ we describe the experiment by a k-dimensional
random variable X= (X$_{1}$,..,X$_{k}$), \ \ \ a common probability space S
and some joint probability density function f$_{X_{1}X_{2}..X_{k}}$ (x$_{1}$%
,..x$_{k}$) . From the joint probability density function one can obtain
various conditional probabilities and by integration over k-1 variables one
obtains k marginal probability density functions f$_{X_{i}}$(x$_{i}$)
describing \textit{k} different random experiments each performed to measure
only one random variable X$_{i}$ and neglecting all the others . In this
case we say that f$_{X_{i}}$ (x$_{i\text{ }}$) were obtained by
conditionalization from a unique probability space S. In general if the
random variables X$_{i}$ are dependent (correlated)

f$_{X_{1}X_{2}..X_{k}}$ (x$_{1}$,..x$_{k}$)$\neq $f$_{X_{1}}$(x$_{1}$) f$%
_{X_{2}}$(x$_{_{2}}$)...f$_{X_{k}}$(x$_{k}$) \ \ \ \ \ \ \ \ \ \ \ \ \ \ \ \
\ \ \ \ \ \ \ \ \ \ \ \ \ \ \ \ \ \ \ \ \ \ \ \ \ \ \ \ \ \ \ \ (2)

Each spin polarization correlation experiment (A,B) is defined by two
macroscopic orientation vectors\ \textbf{A} \ and\textbf{\ B} being some
average orientation vectors of the realistic polarizer. A polarizer\textit{\
A} is defined by a probability distribution d$\rho _{A}(a)$ , where a are
the microscopic direction vectors, a$\in O_{A}=\left\{ a\in S^{(2)};\left|
1-a\cdot \mathbf{A}\right| \leq \varepsilon _{A}\right\} .$ Similarly a
polarizer B is defined by d$\rho (b).$ \ \ The probability p(A,B) that a \
particle 1 passes through the polarizer A and a particle 2 , correlated with
the particle 1 passes through a polarizer B is given by

p(A,B)= $\ \stackunder{O_{A}}{\int \text{ }}\stackunder{O_{B}}{\int }%
p_{12}(a,b)$d$\rho _{A}(a)$d$\rho (b)$ \ \ \ \ \ \ \ \ \ \ \ \ \ \ \ \ \ \ \
\ \ \ \ \ \ \ \ \ \ \ \ \ \ \ \ \ \ \ \ \ \ \ \ \ \ \ \ \ \ \ \ (3)

where $p_{12}(a,b)$ is a probability density\textbf{\ }function given by QM :%
$p_{12}(a,b)$= $\frac{1}{2}\sin ^{2}(\theta _{ab}/2).$In the reference $%
\left[ 13\right] $ we used a slightly different but consistent notation. It
is impossible to perform different experiments(A,B) simultaneously on the
same couple of the particles therefore it does not seem possible to use a
unique probability space S and to obtain,by conditionalization, the
probabilities p(A,B) for all such experiments.

Let us for example analyze a model used by Clauser and Horne $\left[ 37%
\right] $to prove their inequalities:

p(A,B)=$\stackunder{\Lambda }{\int }p_{1}(\lambda ,A)$ $p_{2}(\lambda
,B)d\rho (\lambda )$ \ \ \ \ \ \ \ \ \ \ \ \ \ \ \ \ \ \ \ \ \ \ \ \ \ \ \ \
\ \ \ \ \ \ \ \ \ \ \ \ \ \ \ \ \ \ \ \ \ \ \ \ \ \ \ \ \ (4)\ 

where $p_{1}(\lambda ,A)$ and $p_{2}(\lambda ,B)$ are the probabilities of
detecting component 1 and component 2 respectively , given the state $%
\lambda $ of the composite system .

We see from (4) that a state $\lambda $ is determined by all the values of
strictly correlated spin projections of two components for all possible
orientations of the polarizers A and B. The polarizers are not perfect
therefore the detection probabilities have been introduced. Therefore it is
assumed in the model that even before the detection each component has well
defined spin projection in all directions. Therefore a model is using a
single probability space $\Lambda $ and obtains the predictions on the
probabilities p(A,B) measured in different experiments by
conditionalization. As we told the same assumption was used in all other
proofs of Bell's theorem. Explicit description of states $\lambda $ by the
values of spin projections is also clearly seen in Wigner's proof$\left[ 38%
\right] $.

As we told the experiments (A,B) are mutually exclusive there is no
justification for using such models.

\bigskip

If we try to prove the Bell's inequalities by comparing only the
experimental runs of different experiments we can not do it without some
additional and questionable assumptions.

Let us simplify the argument we gave in$\left[ 14\right] .$We want to
estimate a value of the spin expectation function E$_{AB}$ for an experiment
(A,B) . We have to perform several runs of the length N and find the value
of the empirical spin expectation function r$_{N}$(A,B) for each run and
after to find E$_{AB}$ by averaging over various runs. Let us associate with
each member of a pair a spin function s$_{1}$(x) or s$_{2}$(x), taking the
values 1 or -1,\ on the unit sphere S$^{(2)}$ (representing the orientation
vectors of various polarizers) .We assume also that s$_{1}$(x) =- s$_{2}$%
(x)= s(x) for all vectors x$\in $S$^{(2)}$. We saw in (3) that the
macroscopic directions \textbf{A} and\textbf{\ B} were not sharp therefore
in each particular run we might have different direction vectors (a,b)
representing them.. If for the simplicity we neglect this possibility, we
get:\bigskip

r$_{N}$(A,B)= $\ -\frac{1\text{ \ }}{N}\stackunder{i}{\sum }$ s$_{i}$(%
\textbf{A})s$_{i}$(\textbf{B}) \ \ \ \ \ \ \ \ \ \ \ \ \ \ \ \ \ \ \ \ \ \ \
\ \ \ \ \ \ \ \ \ \ \ \ \ \ \ \ \ \ \ \ \ \ \ \ \ \ \ \ \ \ \ \ \ \ \ \ \ \
(5)

where N functions s$_{i}$ are drawn from some uncountable set of \ \ spin
functions F$_{0}.$

If we consider a particular run of the same length from the experiment (A,C)
we get

\bigskip r$_{N}$(A,C)= $\ -\frac{1\text{ \ }}{N}\stackunder{j}{\sum }$ s$%
_{j} $(\textbf{A})s$_{j}$(\textbf{C}) \ \ \ \ \ \ \ \ \ \ \ \ \ \ \ \ \ \ \
\ \ \ \ \ \ \ \ \ \ \ \ \ \ \ \ \ \ \ \ \ \ \ \ \ \ \ \ \ \ \ \ \ \ \ \ \ \
\ \ \ (6)

where N functions s$_{j}$ are drawn from the same uncountable set of \ spin
functions F$_{0}.$

A probability that we have the same sets of spin functions in both
experimental runs is equal to zero. Therefore in general we have completely
distinct sets of functions in (5) and (6). and we are unable to prove the
Bell's theorem make by using r$_{N}$(A,B)-r$_{N}$(A,C). If we used the same
sets of spin functions in the runs from the different experiments then we
could replace (6) by (7)

\bigskip r$_{N}$(A,C)=$-\frac{1\text{ \ }}{N}\stackunder{i}{\sum }$ s$_{i}$(%
\textbf{A})s$_{i}$(\textbf{C}) \ \ \ \ \ \ \ \ \ \ \ \ \ \ \ \ \ \ \ \ \ \ \
\ \ \ \ \ \ \ \ \ \ \ \ \ \ \ \ \ \ \ \ \ \ \ \ \ \ \ \ \ \ \ \ \ \ \ \ \ \
\ \ (7)

and we could easily reproduce the Bell's proof \ finding \ the similar
inequalities. However the formula is counter-factual and does not represent
the experimental data.

Let us notice the act of passage of the i-th particle through a given
polarizer\ A depends in a complicated way on its interaction with this
polarizer. Therefore we should not consider a spin function as describing a
state of a particle independent of its interaction with A.. The spin
functions s$_{i}$ n the (5) and (6) resume the interactions of the
subsequent particles with the polarizers in a particular experiment.
Therefore if we want to be rigorous we should replace (5) by (8).

r$_{N}$(A,B)= $\ -\frac{1\text{ \ }}{N}\stackunder{i}{\sum }$ s$_{i,\mathbf{A%
}}$(a$_{i}$)s$_{i,\mathbf{B}}$(b$_{i}$) \ \ \ \ \ \ \ \ \ \ \ \ \ \ \ \ \ \
\ \ \ \ \ \ \ \ \ \ \ \ \ \ \ \ \ \ \ \ \ \ \ \ \ \ \ \ \ \ \ \ \ \ \ \ \ \
(8)

where a$_{i}\in O_{A}$ and b$_{i}\in O_{B}.$ \ If we use the formula (8)
there is no possibility of proving Bell's theorem . This formula is
consistent with the probabilistic model (3) and with the contextual
character of observables.

\bigskip

A particular trivial, but artificial, example when a common probability
space S could have been used is a case when we have \textit{4 }independent
random variables X$_{1}$,X'$_{1},$X$_{2}$,X'$_{2}$ each described by its
probability density function. If all these variables have only the values $%
\pm 1$ a proof of Bell's inequalities is extremely easy . The ''spin''
expectation function E$_{X_{1}X_{2}}$ is a product of expectation values of X%
$_{1}$ and X$_{2}$ : E$_{X_{1}X_{2}}=\left\langle X_{1}\right\rangle
\left\langle X_{2}\right\rangle $ and we immediately get

$\bigskip $

$\left| \left\langle X_{1}\right\rangle \left\langle X_{2}\right\rangle
-\left\langle X_{1}\right\rangle \left\langle X_{2}^{\prime }\right\rangle
\right| $ + $\left| \left\langle X_{1}^{\prime }\right\rangle \left\langle
X_{2}\right\rangle +\left\langle X_{1}^{\prime }\right\rangle \left\langle
X_{2}^{\prime }\right\rangle \right| \leq \left| \left\langle X_{1}^{\prime
}\right\rangle -\left\langle X_{2}^{\prime }\right\rangle \right| +\left|
\left\langle X_{1}^{\prime }\right\rangle +\left\langle X_{2}^{\prime
}\right\rangle \right| \leq 2$ \ \ \ (9)

\bigskip

Of course if we assume the independence there are no correlations. The
statistical independence is equivalent \ to separability of the statistical
operator used recently by Kr\"{u}ger in his proofs of Bell's inequalities in 
$\left[ 39\right] $ and in unpublished paper presented at TH 2002.

We find all \ these arguments very convincing but it is well known that a
single picture is better than thousand words. This picture was given by the
computer pseudo-random experiments of Accardi et al.$\left[ 22\right] $
which violate Bell's inequalities giving an example of the family of random
experiments which cannot be described using a unique probability space S.

Therefore the violation of Bell's inequalities found in SPCE proves\ that
the probabilistic model used by \ LRHV is inappropriate but it tells nothing
about completeness of QM or about the impossibility of causal
''microscopic'' explanation of quantum experiments.

Let us examine the ''microscopic''description of the classical experiment E3
which we discussed in one of the preceding sections. We see that this
description does not depend on observed results B or R, it depends on other
physical properties of the coin , on the mechanical properties of the device
and even on the properties of \ the ambient air. All these ''hidden''
parameters explain in very complicated way the observed events. Similarly a
pilot wave models of de Broglie$\left[ 40\right] $, Bohm$\left[ 41\right] $
and Vigier$\left[ 42\right] $ reproduce in a complicated way some of the
quantum mechanical predictions. We like a remark by Tartalia $\left[ 43%
\right] $ that objects in the quantum world are like programmed machines
capable of different behaviors according to the physical conditions locally
triggering them.

Another important implication of (3) and (8) is that the observable value of
the spin projection characterizes only the whole beam of the particles which
passed through a given \ polarizer A. Nearly 100\% of the particles of this
beam pass through a subsequent polarizer A, but we have no prediction
concerning any individual particle from the beam. Therefore in SPCE \ \
p(A,A') $\neq 1$\ and we have no strict spin anti-correlations between the
members of \ each pair$\left[ 13\right] $.

\subsection{ 5.Bertrand's paradox.}

Many probabilists in 19th century believed that for each random experiment
there exists a probability distribution which may be determined only by
combinatorial or geometric considerations .In 1889 Bertrand showed$\left[ 18%
\right] $ that the various equally good mathematical arguments, in case of
the continuous random variables, may lead to completely different
predictions on the probabilities. He considered two concentric circles on a
plane with radii R and R/2, respectively. He showed that there are different
possible answers to a question: '' What is the probability P that a chord of
the bigger circle chosen at random cuts the smaller one at least in one
point?. The various answers are$\left[ 14\right] $: if we divide the
ensemble \ of all chords into sub-ensembles of parallel chords, we find P=
1/2. If we consider the sub-ensembles having the same beginning , we find
P=1/3 . Finally if we look for the midpoints of the chords lying in the
small circle , we find P=1/4 . A solution of the paradox is simple: the
different values of P correspond to the different random experiments which
may performed in order to find the experimental answer to the Bertrand's
question. Thus the probabilities have only a precise meaning if the random
experiments used for their estimation are specified.

Let us suppose now that we have a straight stick of the length 2R. We draw
two concentric circles on the horizontal platform and we construct three
machines M1, M2 and M3 working according to appropriate different
pseudo-random protocols corresponding to different reasonings presented
above..

In the first experiment we insert a stick into the machine M1 which picks up
a ''point'' Q on the large circle, then follows the diameter of the circle
arriving to the point Q$_{1}$ located in the randomly chosen distance r from
the point Q (0 $\leq $ r $\leq $ 2R). Next M1 places the stick
perpendicularly to the diameter joining Q and Q$_{1}$ with the midpoint of a
stick coinciding with the point Q$_{1}.$ \ If the stick touches in at least
one point the smaller circle we may say that the value of the random
variable X is equal to 1 otherwise it's value is -1. After many repetitions
of this experiment we find the probability p(X=1)=1/2.

Similarly in the second experiment we can obtain the probability p(Y=1)=1/3
and in the third p(Z=1)=1/4. It is feasible but unreasonable and artificial
to introduce a unique probability space S and the joint probability
distribution of \ X,Y and Z in order to deduce, by conditionalization., the
probability distributions of our three random experiments.

\subsection{6.Conclusions}

The experimental tests of Bell's theorem can neither confirm the
completeness of QM nor to prove that the only TSP models able to give a
''microscopic'' description of the SPCE have to violate Einstenian causality.

A question whether a statistical description provided by the quantum theory
gives a complete description of the experimental data\textbf{\ }is fully
justified. This question about completeness can not be answered \ by proving
a mathematical theorem or by constructing ad hoc TSP model reproducing the
quantum predictions.

This question of the completeness of quantum theory can be only answered by
a detailed analysis of the time- series of the experimental results which
can be done with the help of the purity tests which were proposed many years
ago and never done.

If the deviations from the randomness were detected and some new
regularities found , the standard statistical description given by the
quantum theory should be completed by a description using probably the ideas
of the stochastic processes. In some sense this change of the description
has already been made in the stochastic approaches used to explain various
phenomena involving trapped atoms, ions and molecules. In these approaches
the wave function obeys a Schr\H{o}dinger equation with an effective
Hamiltonian separated by quantum jumps occurring at random times. The purity
tests could be also used to check these new stochastic models which assume
without checking the ergodicity of the observed time-series. The question of
the completeness formulated in this way is independent of the existence or
non existence of a detailed ''microscopic'' description of the phenomena
presenting this particular stochastic behavior.

From Bertrand's paradox we learned that we should not talk about the
probabilities without referring to the random experiments used to determine
them \ Therefore the quantum theory providing the predictions for various
probabilities should not loose its contact with the experiments it wants to
describe.

If one forgets that the quantum theory does not give any ''microscopic''
images but it provides only the mathematical algorithms able to describe the
statistical regularities observed in the data one is tempted to create
incorrect mental ''microscopic'' images which lead to false paradoxes and to
speculations which seem to be a pure science fiction.

The quantum observables are contextual what means that their values are not
the attributes of the individual members of the quantum ensemble but they
give only the information about the possible interactions of \ the whole
ensemble with the measuring devices. If the ensemble is pure one can speak
about the probabilistic information pertinent to the interaction of each
individual system with the measuring device. To be able to do this one must
check the purity of the ensemble using the purity tests.

There is no strict anti-correlations of two time-series of the results of
measurements performed on two members of EPR pair in the SPCE thus these two
time series may not be used in quantum cryptography to assure that Bob and
Alice use the same secret key.

The purity tests are important and relatively simple , the data are
available. We hope that this paper will convince some experimentalists to do
them.

\subsection{References}

\begin{enumerate}
\item  N.Bohr,Essays 1958-1962 on Atomic Physics and Human Knowledge( Wiley,
New York,1963)p.60.

\item  A.Einstein., B.Podolsky and N. Rosen, Phys.Rev. 47(1935) 777.

\item  L.E.Ballentine, Rev.Mod.Phys. 42(1970)358.

\item  F.J.Belinfante, A survey of hidden variable theories (Pergamon, New
York, 1973).

\item  J.S.Bell, Physics 1(1965)195..

\item  J.F.Clauser, M.A.Horne,A.Shiminy and R.A. Holt,
Phys.Rev.Lett.23(1969) 880.

\item  A.Aspect, P.Grangier and G.Roger, Phys.Rev.Lett.47 (1981) 460;
49(1981) 91:

Aspect, J. Dalibard and G.Roger, Phys. Rev.Lett. 49 (1982) 1804.

\item  L.Accardi: Phys.Rep.77(19810 169.

\item  I. Pitovsky, Phys.Rev.Lett.49(1982) 1216 and Phys.Rev.D 27(1983) 2316.

\item  D.Aerts, J.Math.Phys. 27(1986)202

\item  W.de Baere, , Lett .Nuovo Cimento 39(1984) 234; 40(1984) 488.

\item  M.Kupczynski, New test of completeness of quantum mechanics, ICTP
priprint IC/84/242, (1984)

\item  M.Kupczynski, Phys.Lett.A 121(1987) 51.

\item  M.Kupczynski, Phys.Lett.A 121(1987)205.

\item  J.F.Clauser and A.Shimony, Rep.Prog.Phys 41(1978) 1883.

\item  F.E.Schroeck,Jr., Found. Phys. 15(1985) 279.

\item  E.Prugovecki,Stochastic quantum mechanics and quantum space time,

(Reidel,Dordrecht ,1984).

\item  J.Bertrand, Calcul des probabilit\'{e}s( Paris, 1889) p.4

\item  C.H.Bennet and G.Brassard in the Proceedings of IEEE International
Conference on Computers, Systems, and Signal Processing, Bangalore,
India(IEEE, New York, 1984) p.175

\item  C,H.Bennet,G.Brassard and N.D.Mermin, Phys.Rev.Lett.68 (1992) 557.

\item  A.Ekert, Phys.Rev.Lett. 67 (1992) 557.

\item  L.Accardi and M.Regoli: Non-locality and quantum theory:new
experimental evidence, Preprint Volterra(2000), N 420.

\item  L. Accardi, On the EPR paradox and the Bell inequality, Volterra
preprint (1998) N. 350

\item  J.Maddox, Nature (1988) 335.

\item  4) L. Accardi, Massimo Regoli:Locality and Bell's inequality,in:
QP-XIII, Foundations of Probability and Physics, A.Khrennikov(ed.), World
Scientific (2002) 1--28

\item  M.Kupczynski, Phys.Lett. A 116(1986), 417.

\item  G.Birkhoff and J.von Neumann, Ann.Math. 37(1936),827.

\item  B.Mielnik, Commun.Math.Phys. ,9(1968)55.;15(1969),55;37,(1974)221.

\item  M.Kupczynski, Int.J.Theor.phys.791973) 319, reprinted in: Physical
theory as logico-operational structure,ed.
C.A.Hooker(reidel,Dordrecht,1978),p.89

\item  M.Kupczynski,Lett.Nuovo.Cimento 11(1974)134.

\item  M.Kupczynski,Lett.Nuovo.Cimento 11(1974)117.

\item  M.Kupczynski,Lett.Nuovo.Cimento 11(1974)121..

\item  M.Kupczynski,Hadronic J. 9 (1986)215.

\item  M.Kupczynski,Riv.nuovo Cimento 7(1977)215.

\item  M.Kupczynski, Phys.Lett.B 47(1973)244.

\item  M.Kupczynski,Czech.J.Phys.1327(1977)17.

\item  J.Clauser and M.A.Horne, Phys.Rev.D10(1974)526.

\item  E.P.Wigner,Am.J.Phys.38(1970)1003.

\item  T.Kr\"{u}ger, Found.Phys.30,1869(2000)

\item  L.de Broglie,Non linear wave mechanics,(Elsevier,1960)

\item  D.Bohm, Phys.Rev.85,(1952)166 and 180

\item  D.Bohm and J.P.Vigier, Phys.Rev.96(1954)208

\item  A.Tartaglia, Eur.J.Phys.19(1998),307.
\end{enumerate}

\subparagraph{\protect\bigskip}

\end{document}